\documentclass{PoS}
\usepackage{amsmath}
\usepackage{amssymb}
\usepackage{graphicx}
\usepackage{float}
\usepackage{cite}

\newcommand{\sqrtS}{\ensuremath{\sqrt{s}}}

\newcommand{\GeVc}{\ensuremath{\mathrm{GeV}\kern-0.05em/\kern-0.02em c}}
\newcommand{\pT}{\ensuremath{p_{\mathrm{T}}}}

\newcommand{\dEdx}{\ensuremath{\mathrm{d}E/\mathrm{d}x }}

\newcommand{\pp}{\ensuremath{\mathrm {p\kern-0.05em p}}}

\newcommand{\K}{\mbox{$\mathrm {K}$}}

\newcommand{\kshort}{\mbox{$\mathrm {K^0_S}$}}

\newcommand{\kstarZ}{\mbox{\K$^{*}$(892)$^{\mathrm{0}}$}}

\newcommand{\kstarch}{\mbox{\K$^{*}$(892)$^{\mathrm{\pm}}$}}
\newcommand{\simplekstarch}{\mbox{\K$^{*\mathrm{\pm}}$}}

\title{First results on \ensuremath{{\rm K^{*}(892)}^{\pm}} resonance production in pp collisions with ALICE at the LHC}

\ShortTitle{\kstarch~ resonance production}

\author{\speaker{Kunal Garg}\thanks{on behalf of the ALICE collaboration}\\
        Department of Physics and Astronomy, University of Catania  and INFN -Sezione di Catania\\
        E-mail: \email{kunal.garg@cern.ch}}

\abstract{The study of strange hadronic resonances in pp collisions contributes to the study of strangeness production in small systems. Measurements in pp collisions constitute a reference for the study in larger colliding systems and provide constraints for tuning QCD-inspired event generators. Since the lifetimes of short-lived resonances such as \kstarch~($\tau \sim 4$ fm/\textit{c}) are comparable with the lifetime of the fireball produced in heavy-ion collisions, regeneration and rescattering effects can modify the measured yield, especially at low transverse momentum. \\
\\
The first results for the \kstarch~resonance obtained in inelastic pp collisions at \sqrtS~= 5.02, 8, and 13 TeV  will be shown. The \kstarch~has been measured at mid-rapidity via its hadronic decay channel $\simplekstarch~\rightarrow \kshort~+\ensuremath{\pi^{\pm}}$, with the ALICE detector.  In particular, the transverse momentum (\pT) spectrum, integrated yields, and ratio to stable hadrons will be presented. The results from \kstarch~are compared to the \kstarZ. The \kstarch~results are compared with \kstarch~predictions from various theoretical models. Particularly interesting is a comparison of the \pT~spectra obtained at different energies that shows hardening of the spectra with increasing energy of the collisions. }

\FullConference{Sixth Annual Conference on Large Hadron Collider Physics (LHCP2018)\\
                4-9 June 2018\\
                Bologna, Italy}

\begin{document}

\section{Introduction}
The measurement of the production of strange resonances in ultrarelativistic proton-proton collisions permits characterising the global properties of the collisions and probing strangeness production. Furthermore, it helps in understanding hadron production processes and in improving the description of hadronisation of strange particles in event generators such as PYTHIA \cite{1,2}, PHOJET \cite{3}, EPOS-LHC \cite{4}. Additionally, the measurement of resonances over a large transverse momentum range and at different collision energies permits probing the perturbative (hard)  and the non-perturbative (soft) QCD processes.
\newline
Hadronic resonances are also important probes of the hadronic phase of the medium formed during the heavy-ion collisions.  \kstarch~is a strange vector meson with a lifetime of $\sim$ 4 fm/c, comparable to the one of the fireball produced in heavy-ion collisions. Thus it is expected that \kstarch~mesons will go through the re-generation and re-scattering processes occurring during the hadronic phase. Measurements in pp collisions constitute a reference for the study in larger colliding systems.
\newline
This paper focus on the \kstarch~production and on its collision energy dependence, comparing results obtained with the ALICE detector in pp collisions at \sqrtS~= 5.02, 8 and 13 TeV at the LHC. In particular, \pT~spectra at different energies as well as \pT-integrated particle ratios to long-lived kaons are compared. In this paper, we address the questions if there is any collision energy dependence of \kstarch~production in pp collision, and how the theoretical predictions compare with data for the three collision energies.

\section{\kstarch~ meson reconstruction and \pT~ spectra}
The \kstarch~is reconstructed at mid-rapidity (|$y$| < 0.5) via its hadronic decay channels i.e. $\pi^{\pm} + \kshort$, where \kshort~is reconstructed by its weak decay $\kshort~\rightarrow \pi^{+} + \pi^{-} $ with a branching ratio of 69.2$\%$. The primary pions are identified through their energy loss \dEdx~in the Time Projection Chamber (TPC), while \kshort~is identified by applying topological cuts on the daughter tracks. The yields of \kstarch~in different \pT~intervals are obtained by subtracting the combinatorial background estimated with an event-mixing technique from the invariant mass distribution. The obtained invariant mass distribution is fitted with a non-relativistic Breit-Wigner function for the signal summed to the residual background parametrised as in \cite{5}:
\begin{eqnarray}
F_{BG}(M_{\rm{K\pi}}) = [M_{\rm{K\pi}} - (m_{\rm{\pi}} + m_{\rm{K}})]^n exp(A + B M_{\rm{K\pi}} + C M_{\rm{K\pi}}^2)
\label{eqn1}
\end{eqnarray}

\noindent where $M_{\rm{K\pi}}$ is the variable representing the mass of the resonance, $m_{\rm{\pi}}$ is the mass of the pion, $m_{\rm{K}}$ is the mass of the \kshort, and A, B, and C are constants. In different \pT~intervals, raw yields are obtained from the residual background-subtracted signal distributions. The inelastic yield is then estimated normalising to the number of minimum bias events and correcting for the detector efficiency times acceptance, the branching ratio, the signal loss correction which accounts for loss of signal due the use of kINT7 trigger (logical AND between signals from V$0$A and V$0$C detectors), the vertex correction factor which accounts for the signal loss introduced by the requirement that a primary vertex be reconstructed, and the inelastic normalisation factor that takes into account the trigger efficiency.

\section{Results and Discussion}

\begin{figure}[h]
\includegraphics[width= 0.45\textwidth]{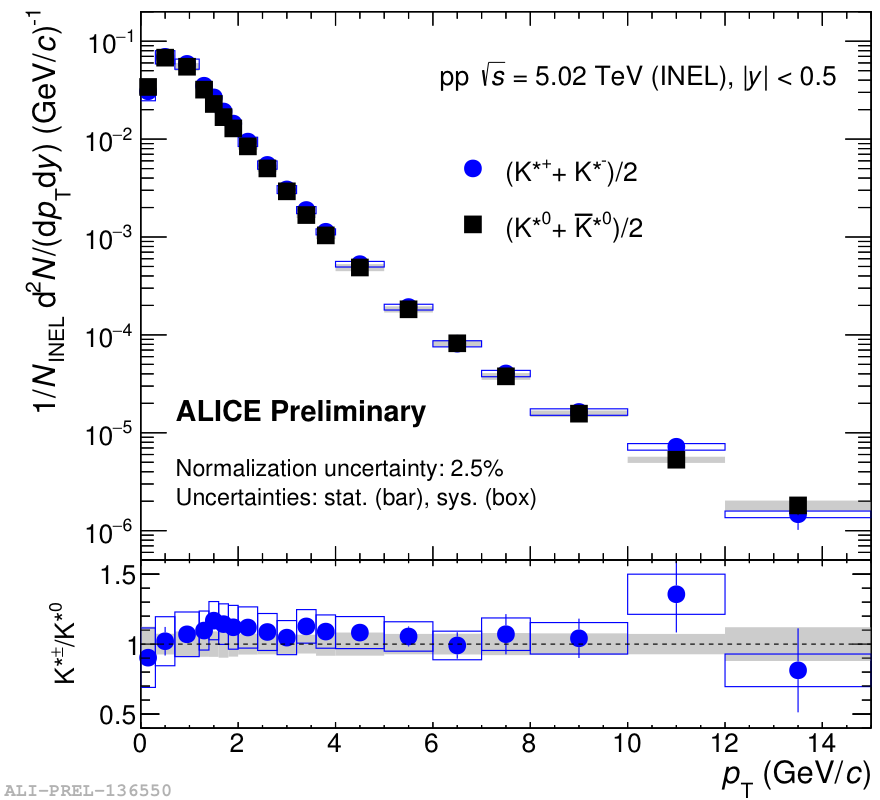}
\includegraphics[width= 0.45\textwidth]{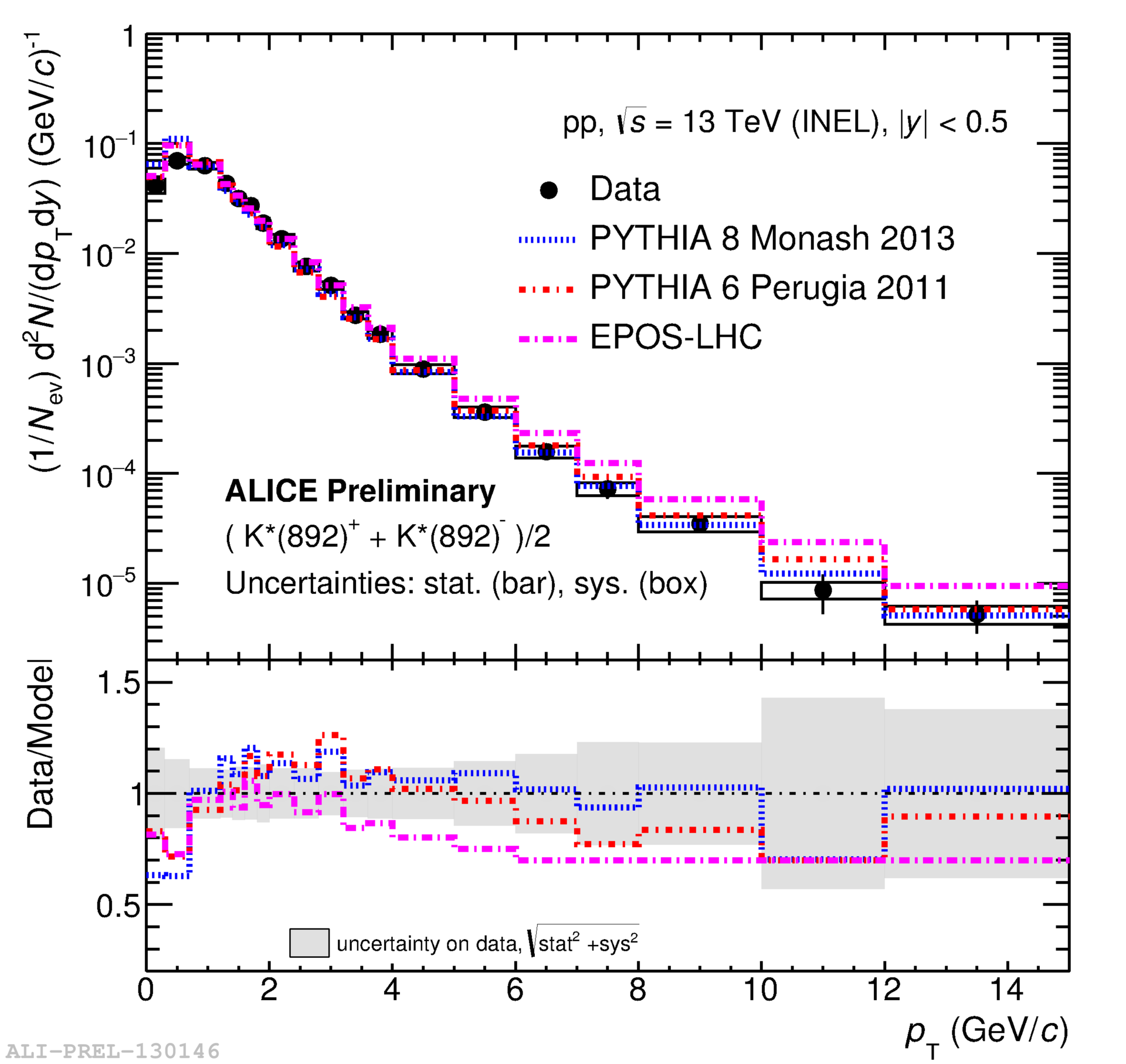}
\caption{Left: Comparison between neutral and charged $\rm{K^{*}}$ \pT~spectra for pp collisions at \sqrtS~= 5.02 TeV Right: \kstarch~ \pT~ Inelastic (INEL) spectra at \sqrtS~= 13 TeV compared with different models.}
\label{fig2} 
\end{figure}

The left panel of Fig. \mbox{\ref{fig2}} shows the charged K* spectrum obtained at 5.02 TeV compared with the \kstarZ~spectrum measured at the same energy. It is evident that the charged and neutral K* productions are equal within the estimated uncertainties. In the right panel of Fig. \mbox{\ref{fig2}} the comparison of the \kstarch~\pT~spectrum measured  at 13 TeV with the predictions of different event generators (PYTHIA8 - Monash 2013\cite{1}, PYTHIA6 - Perugia 2011 \cite{2} and EPOS-LHC \cite{4}) is shown. All the models overestimate the production at low \pT~(\pT~< 1 \GeVc), while EPOS-LHC overestimates the production at high \pT~too.
\newline

\begin{figure}[h]
\includegraphics[width= 0.45\textwidth]{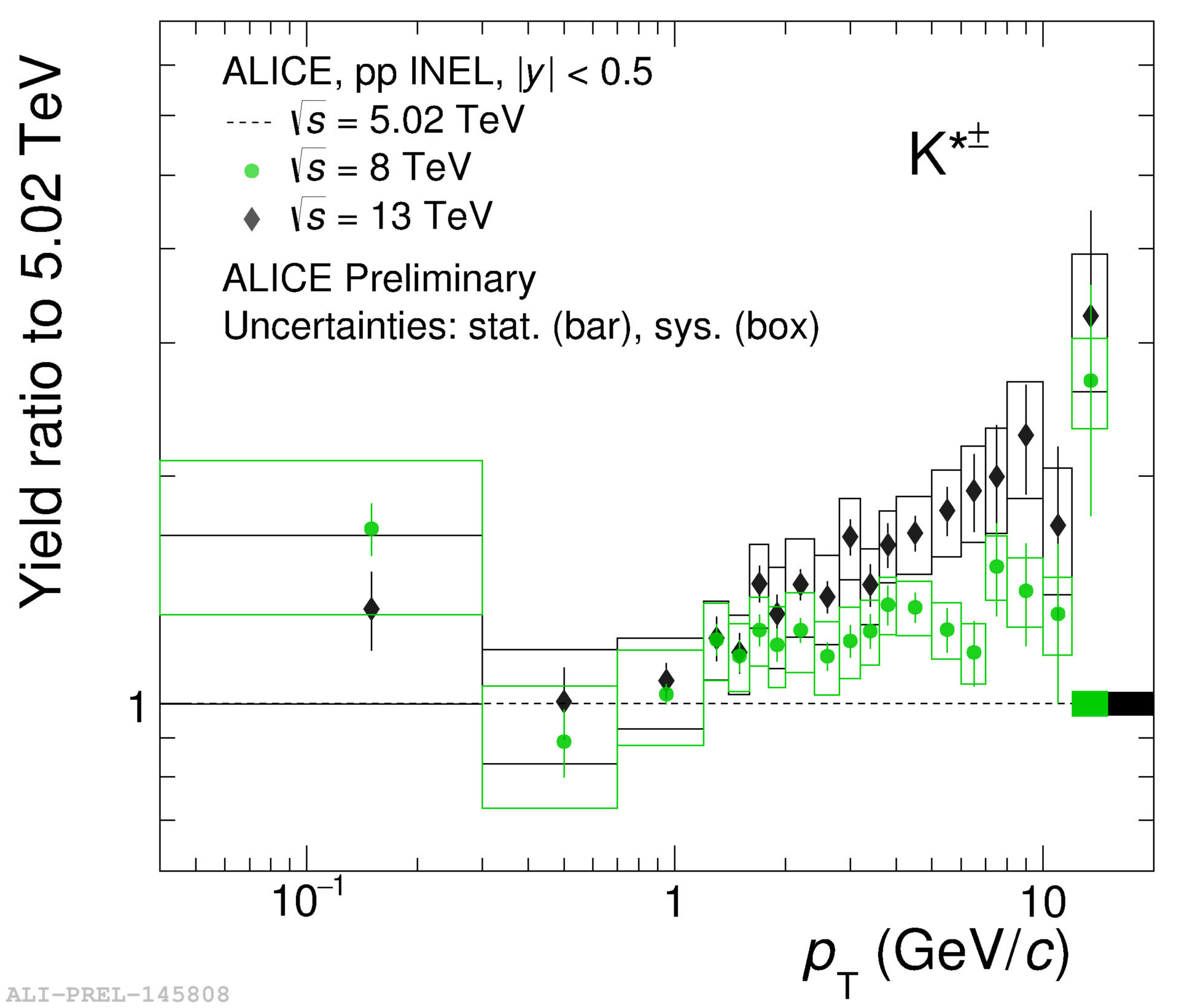}
\includegraphics[width= 0.45\textwidth]{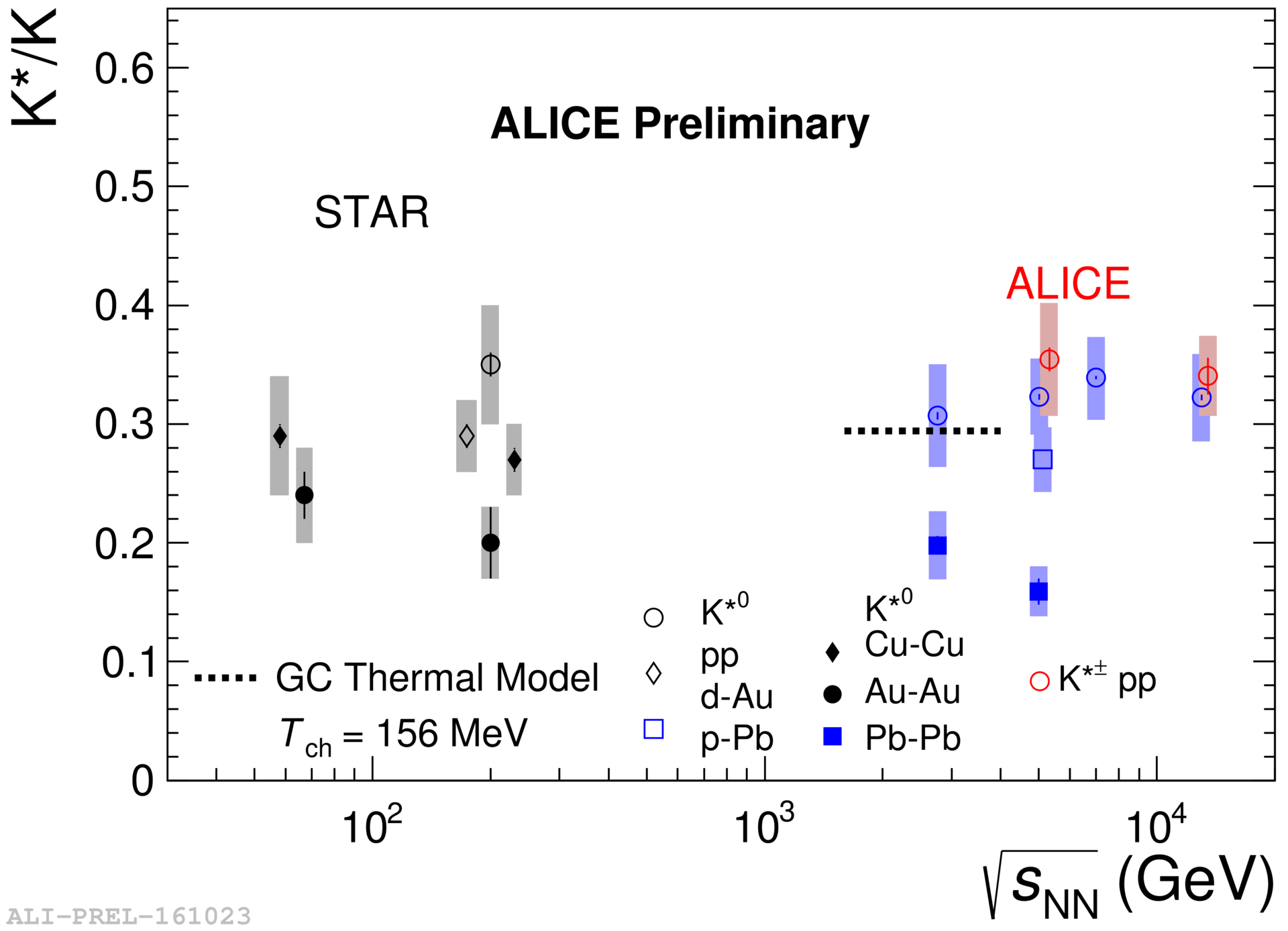}
\caption{Left: Ratios of \pT~ spectra of \kstarch~ in inelastic pp collisions at various centre-of-mass energies to the spectrum obtained in pp \sqrtS~= 5.02 TeV. Right: $\rm K^{*}/K$ ratio vs. collision energy. References to STAR and ALICE data can be found in \cite{6,7,8,9}}
\label{fig1} 
\end{figure}

Figure \mbox{\ref{fig1}} shows ratios of \kstarch~\pT~spectra in inelastic pp collisions at \sqrtS~= 8 and 13 TeV to the spectrum obtained at \sqrtS~= 5.02 TeV. These ratios indicate that the high-\pT~part of the spectra increases as a function of collision energy while the bulk production at low \pT~does not strongly depend on the collision energy. The right panel of the figure shows the ratio of the $\rm K^{*}$ (neutral and charged) to the stable kaons in various collision systems \cite{6,7,8,9}. No significant energy dependence is observed in pp collisions. The suppression of $\rm K^{*0}$/K ratios in central heavy-ion collisions with respect to pp collisions could be indicative of dominant elastic re-scattering effects in the hadronic phase. A similar behaviour is also observed for other resonances and stable hadrons \cite{10,11}. $\rm K^{*\pm}$/K ratio at 5.02 and 13 TeV is equal to $\rm K^{*0}$/K ratio within the uncertainties. \\

\section{Summary}
\kstarch~production is studied as a function of the collision energy. For \pT~> 1\GeVc~a hardening of the \pT~spectra is observed with increasing collision energy. Yield and spectra for the \kstarch~are found to be equal to the \kstarZ. The $\rm K^{*}$/K ratio remains fairly constant in a large range of pp collision energies from RHIC to LHC, while a suppression is observed in central heavy-ion collisions indicating that re-scattering dominates over regeneration during the fireball expansion. PYTHIA8 - Monash 2013 and PYTHIA6 - Perugia 2011 have similar predictions and for \pT~> 1 \GeVc~give a reasonable estimate of the measured inelastic \pT~spectrum at \sqrtS~= 13 TeV.

\end{document}